\documentclass[nofootinbib,aps,pre,twocolumn,showpacs,superscriptaddress,groupedaddress]{revtex4-1} \pdfoutput=1
\usepackage{graphicx} 
\usepackage{amsmath}
\usepackage{todonotes}
\usepackage{graphicx}
\usepackage{natbib}
\usepackage{color}

\usepackage{amsthm, amsmath, amssymb}
\usepackage[loose,nice]{units} 

\def\p{\mathbf{p}}

\def\bw{\boldsymbol{w}}

\begin{document}
\title{Regular Regimes of the Three Body Harmonic System}
\date{\today}
\author{Ori Saporta Katz} \affiliation{Department of Applied Mathematics,
Weizmann Institute of Science, Rehovot 76100, Isreal} \author{Efi
Efrati} \email{efi.efrati@weizmann.ac.il} \affiliation{Department of
Physics of Complex Systems, Weizmann Institute of Science, Rehovot
76100, Israel}

\begin{abstract} 
The symmetric harmonic three-mass system with finite rest lengths, despite its apparent simplicity, displays a wide array of interesting dynamics for different energy values. At low energy the system shows  regular behavior that produces a deformation-induced rotation with a constant averaged angular velocity. As the energy is increased this behavior makes way to a chaotic regime with rotational behavior statistically resembling L\'evy walks and random walks. At high enough energies, where the rest lengths become negligible, the chaotic signature vanishes and the system returns to regularity, with a single dominant frequency. The transition to and from chaos, as well as the anomalous power law statistics measured for the angular displacement of the harmonic three mass system are largely governed by the structure of regular solutions of this mixed Hamiltonian system. Thus a deeper understating  of the system's irregular behavior requires mapping out its regular solutions. 

In this work we provide a comprehensive analysis of the system's regular regimes of motion, using perturbative methods to derive analytical expressions of the system as almost-integrable in its low- and high-energy extremes. The compatibility of this description with the full system is shown numerically. In the low-energy regime, the Birkhoff normal form method is utilized to circumvent the low-order 1:1 resonance of the system, and the conditions for Kolmogorov-Arnold-Moser theory are shown to hold. The integrable approximations provide the back-bone structure around which the behavior of the full non-linear system is organized, and provide a pathway to understanding the origin of the power-law statistics measured in the system.
\end{abstract} \maketitle

\section{introduction}

\begin{figure}
\includegraphics[width=0.7\linewidth]{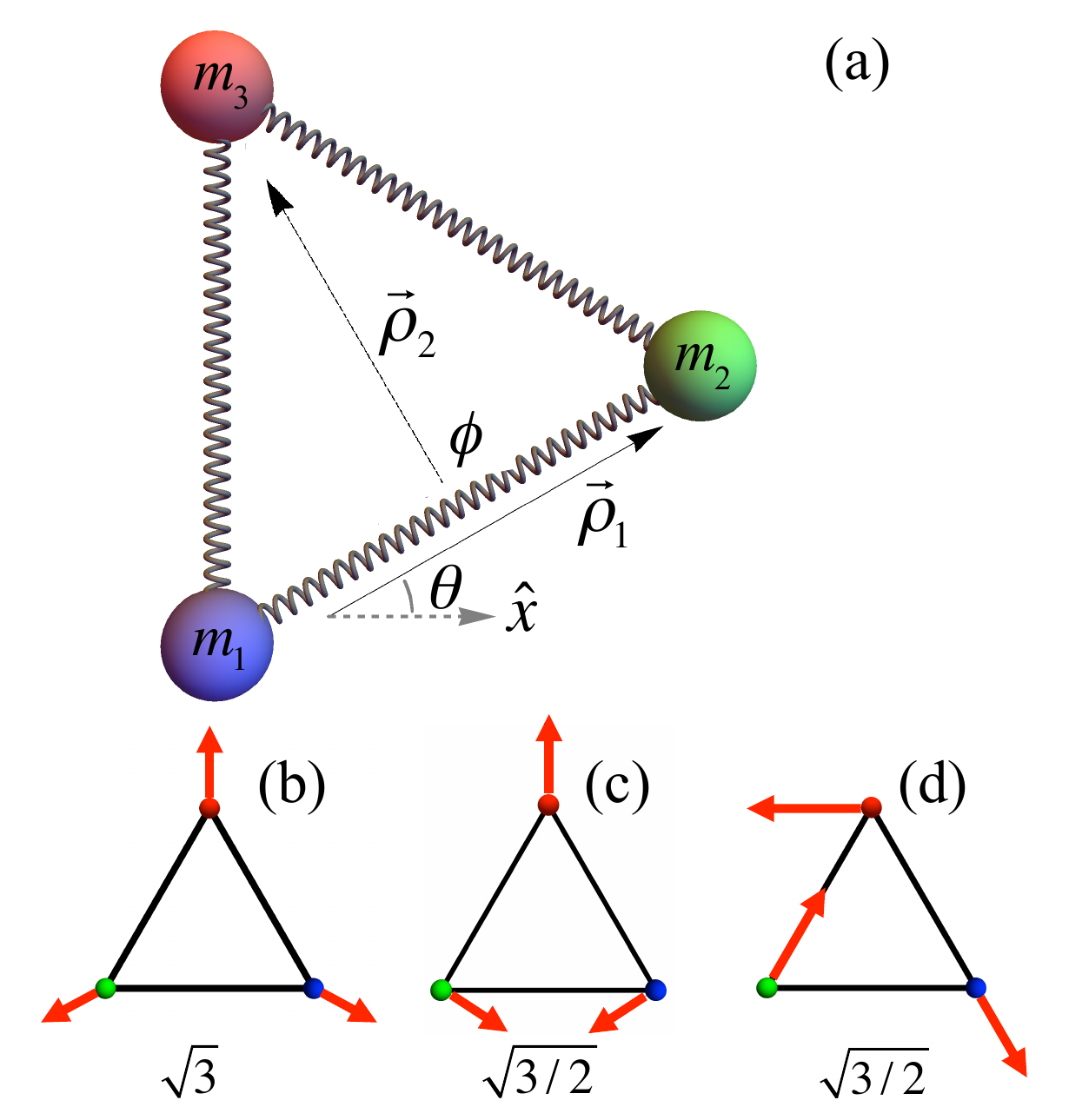}
\caption{(a) The symmetric harmonic 3-mass system, with equal spring rest lengths $L$, equal spring constants $k$ and equal masses $m$. $\vec{\rho}_1$ and $\vec{\rho}_2$ are the mass-weighted Jacobi coordinates, $\phi$ is the angle between them and $\theta$ is the orientation variable of the triangle. (b), (c) and (d) are the system's normal modes and corresponding frequencies, commonly known as the symmetric stretch, isometric bend and asymmetric bend, respectively.}
\label{fig:3masssystem}
\end{figure}

Recently, the harmonic three-mass system with finite rest lengths (Fig. \ref{fig:3masssystem}), was studied and its statistical behavior analyzed \cite{us}. This deceptively simple system was shown to display a rich variety of dynamics due to geometric non-linearities induced by the finite rest lengths of the springs, which render the system dynamically mixed. For different energies and initial conditions, the system exhibits constant deformation-induced rotation with zero angular momentum and random walk of the orientation angle, among other phenomena. Perhaps the most surprising dynamical feature exhibited by the system is the L\'evy-walk regime: for a continuous range of energies, the orientation of the system as a rotating triangle performs bouts of constant average velocity, switching directions with a power-law distribution, fitting the L\'evy walk model \cite{Geisel1987, Geisel1992, Zaburdaev2015}. The anomalous exponent attributed to this dynamics seems to interpolate smoothly between the value of $2$, signifying coherent ballistic behavior, and $1$, signifying regular random walk statistics. 

In low-dimensional systems, $d\leq 2$, the emergence of power-law statistics is well-understood, attributed to the breakdown of Kolmogorov-Arnold-Moser (KAM) tori creating partial transport barriers which can be crossed by chaotic trajectories at a slow rate in a phenomenon commonly referred to as sticking or trapping \cite{lange2016mechanism, mackay1984transport, meiss1986markov, cristadoro2008universality, AFM14, alus2017universal}.
However, despite the robustness of this phenomenon \cite{latora1999superdiffusion, cagnetta2015strong, Shlesinger1987, Zaburdaev2015, zaslavsky2002chaos},  a general framework for the origin of power-law statistics in high-dimensional mixed Hamiltonian systems continues to elude current understanding \cite{lange2016mechanism, danieli2017intermittent, das2019power, meiss2015thirty}. In the three-body harmonic system, the coherent bouts creating the power-law statistics strongly resemble their lower-energy regular counterparts, indicating a partial trapping of chaotic trajectories around regular islands for finite times. A quantitative analysis of the dynamical mechanism behind this phenomenon would require a deep understanding of the regular behavior of the non-linear system.

In this work we seek to identify and characterize the regular solutions of the harmonic three body system, complementing the work in \cite{us}, in the extreme low- and high-energy regimes. 
By using a perturbative approach we find integrable approximations of the Hamiltonian and characterize their solutions. We show how presenting the dynamics of the full system in the phase space variables induced by the integrable approximations leads to a simplified picture that allows a clearer interpretation of the chaotic dynamics. Section II presents the system and its interesting dynamics, and provides a brief summary of the results of \cite{us}, as well as some extensions.
 Section III deals with the low-energy regular motion, where energy confines springs to small oscillations. The Birkhoff normal form method is employed in order to obtain a faithful description of the full system as an almost-integrable system, and the conditions for the Kolmogorov-Arnold-Moser (KAM) theory are shown to hold. In Section IV we analyze the high-energy regular motion, where the springs' rest lengths become practically negligible and the system behaves like the harmonic three mass system with vanishing rest lengths (which is quadratic and thus integrable). Section V contains a short summary and a discussion of the outlook of this work.

This analysis sets the stage for a more complete understanding of the behavior observed for intermediate energies, as the phase space structure in the regular regimes is somewhat retained in the anomalous regimes close enough to the regular regimes, and a gradual breaking of this structure results in a continuous transition to, and from, fully chaotic behavior as the energy is raised. 

\section{The Harmonic three mass System}
The Hamiltonian of the planar, fully symmetric three-mass system with non-zero rest lengths is
\begin{equation} 
\begin{aligned}
\mathcal{H}=&\sum_{i=1}^{3}\frac{\p_{i}^{2}}{2m}+\sum_{<ij>}\frac{k}{2}\left
(r_{ij}-L\right)^{2}, 
\label{eq:hamiltonian} 
\end{aligned}
\end{equation} 
where $\boldsymbol{r}_{i} = \left(x_{i},y_{i}\right)$, $\boldsymbol{r}_{ij}=\boldsymbol{r}_{i}-\boldsymbol{r}_{j}$ and $r_{ij}=|\boldsymbol{r}_{ij}|\equiv\sqrt{\boldsymbol{r}_{ij}\cdot\boldsymbol{r}_{ij}}$ for $i,j=1,2,3$. 
The mass $m$, spring constant $k$ and rest length $L$ give rise to a natural time scale $\tau_{s}=\sqrt{m/k}$, and energy scale $E_{s}=\frac{3}{2}kL^{2}$, the energy it takes to contract the system to a point. The parameters we use in simulations and in the following calculations are $L=2$, $k=1$ and $m=1$, giving the typical time $\tau_{s}=1$ and natural energy scale   $E_{s}=6$. Zero-energy equilibrium is achieved when the distances between the masses equal the rest lengths and the masses are at rest.

Conservation of linear and angular momentum reduce the 
12-dimensional phase space of the system to a 6-dimensional phase space. Energy conservation further reduces the dimension of the submanifold of any given trajectory to five. As the internal motions and center-of-mass motion of the system decouple, it is straightforward to eliminate the four center of mass coordinates from the Hamiltonian. The reduction of the two angular momentum degrees of freedom requires Routh reduction \cite{ArnoldDynamical1988} due to the non-holonomic nature of the conservation law, as described in section III. As a result, even when setting the overall angular momentum of the system to zero, the distorting triangle may exhibit deformation-induced rotation, appearing as a manifestation of a relevant geometric phase \cite{Guichardet1984, wilczek1989geometric, Montgomery1993}. Thus, the orientation of the triangle is a non-trivial, history-dependent variable of the system, and serves as a sensitive measurable for the type of dynamics the system follows. 

\begin{figure*}
\includegraphics[scale=0.7]{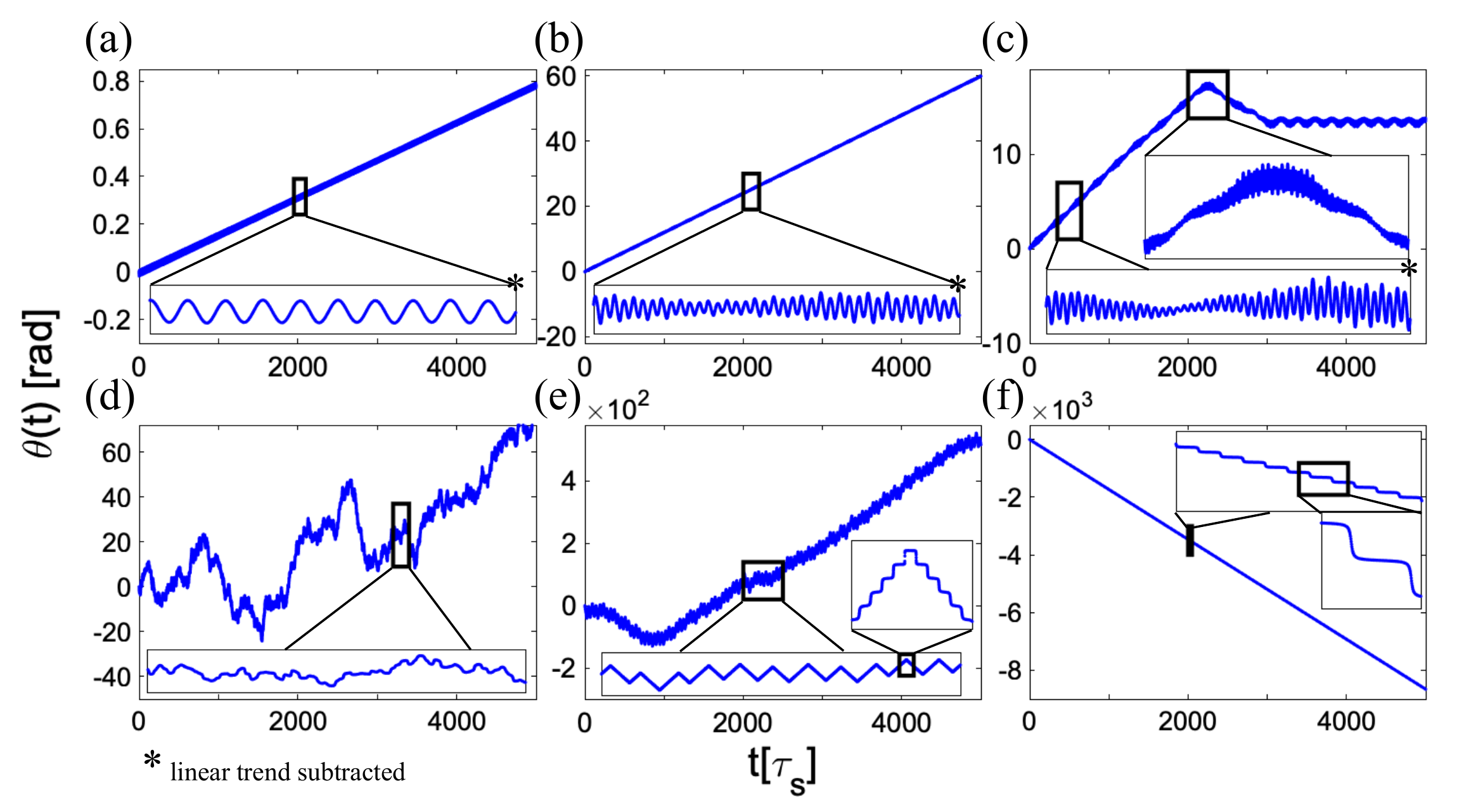}
\caption{Typical angular trajectories of the system for various energies. (a) with $E = 0.005$, and (b) with $E = 0.224$, both exhibit regular behavior for practically infinite times. (a) has a single dominant frequency $\sqrt{3/2}$, which is the twice-degenerate frequency of the linearized reduced system; while (b) shows two dominant frequencies close to the linear frequency resulting in a beating phenomenon. (c) has $E =  0.30$, and is in the L\'evy-walk domain; the trajectory transitions between different seemingly regular trajectories with a power-law distribution. (d) has $E = 1.87$ and exhibits regular diffusion statistics. (e) $E = 770.10$ retains some regularity of motion, and (f) $E = 7.79 \cdot 10^7$ exhibits regular motion corresponding to the linearized high-energy system obtained by setting the rest length to zero.}
\label{fig:typicaltraj}
\end{figure*}

Indeed, the dynamics of the system is incredibly rich: despite the fully harmonic interactions, the non-zero rest lengths of the springs contribute geometrical non-linearities to the system, as can be seen algebraically in the square-root term of the potential energy. This renders the system dynamically mixed, with regions of regular and chaotic dynamics. Reducing our scope to only zero angular momentum configurations, we can characterize the dynamical regime by observing the orientation dynamics. In Fig. \ref{fig:typicaltraj}, the orientation of the triangle is shown for different regimes of motion, and the underlying character of the 6-dimensional dynamics, be it regular, anomalous or chaotic, is apparent through this one-dimensional measurable.
We note that the numerics in Fig. \ref{fig:typicaltraj} and throughout this work have been done using the symplectic integrator provided in \cite{BeronVeraSymplecticNote}, using the symplectic Euler method \cite{Hairer2015}.

\begin{figure}
\includegraphics[scale=0.38]{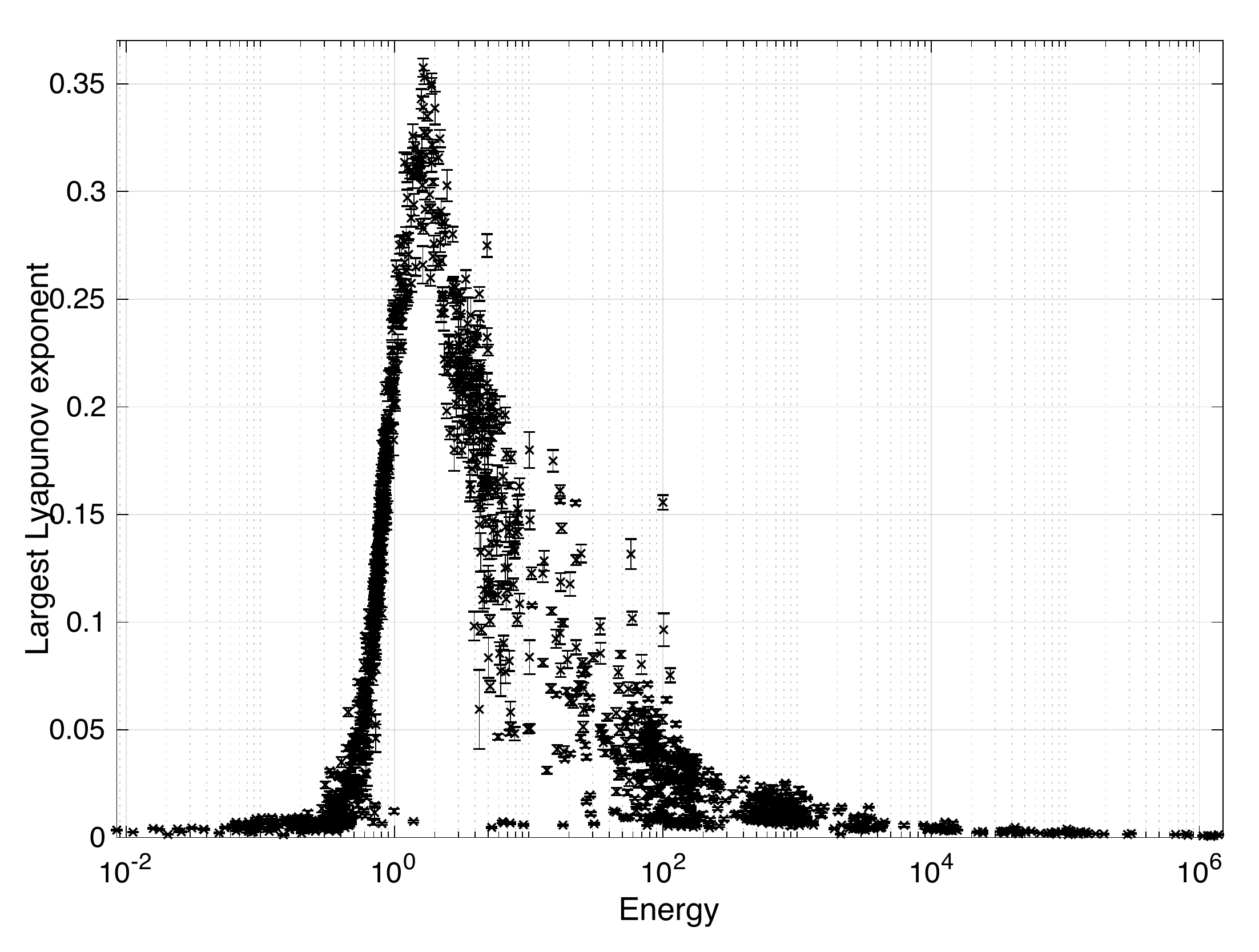}
\caption{Maximal Lyapunov exponents as a function of energy on a log scale, for various random initial conditions with zero angular momentum. At extremely low and high energies, the system behaves regularly and the maximal Lyapunov exponents are zero. In mid-range energies, the system has a chaotic signature with a positive maximal Lyapunov exponent for most initial conditions.
The Lyapunov exponent calculation was done using the method described in \cite{wolf1985determining}, using the Matlab program ``Calculation Lyapunov Exponents for ODE" version 1.0.0.0 by Vasiliy
Govorukhin.}
\label{fig:Lyapunovs}
\end{figure}

A random exploration of different initial conditions shows that for a large portion of trajectories the total energy of the system suffices to describe the statistical quality of the dynamics (see Figure \ref{fig:Lyapunovs}, and \cite{us}), despite the complex structure of the mixed phase space. 
For very low energy values $0<E\ll E_s$ (Fig. 2(a,b)), as well as for very high energy values $E_s \lll E$ (Fig. 2(f)), the system displays stable quasi-periodic regular trajectories with constant averaged deformation-induced rotation rates, and a vanishing Lyapunov exponent.
In the range of energies $E_s/9 \lesssim E \lesssim 2 E_s$ (Fig. 2(d)), orientation trajectories statistically resemble uncorrelated random walks, with a squared mean angular displacement exponent of $1$, and a corresponding positive Lyapunov exponent.

It is in the transition between these three regimes that the exotic dynamics of this system is apparent.
For energy values in the range $E_s/15\lesssim E\lesssim E_s/9$ (Fig. 2(c)), most trajectories exhibit a positive Lyapunov exponent, signifying chaotic dynamics. However, the corresponding squared mean angular displacement exponent is anomalous, transitioning smoothly from the value $2$, corresponding to the ballistic motion characterizing the low energy, to the value $1$, which characterizes the uncorrelated random walks observed for the moderate energy values $E_s/9\lesssim E\lesssim 2 E_s$. The trajectories in this regime display some regularity, following a quasi-periodic trajectory with a constant averaged rotation velocity for a finite time, then transitioning to following a different quasi-periodic trajectory. This ``sticking dynamics" \cite{zaslavsky2000hierarchical}, which results in the emergence of anomalous diffusion, is thus largely determined by the regular quasi-periodic solutions of the Hamiltonian. The transitions themselves between the seemingly quasi-periodic trajectories, while unpredictable, are also related to the underlying phase space structure. Understanding the complex and subtle nature of this dynamics requires a deep understanding of the regular solutions of the system, and the corresponding structure of phase space. 

\section{Low Energy regime}

 In the low-energy regime $E\ll E_s$, trajectories exhibit a regular quasi-periodic motion. This regular motion seems to suggest that a perturbative approach around the system's zero-energy equilibrium would provide a good description of the motion. However, as explained in \cite{us}, linearization in the Cartesian coordinates of Eq. \eqref{eq:hamiltonian} requires breaking the rotational symmetry of the problem by choosing a specific equilibrium position in the plane about which the linearization is performed. This type of linearization, appearing in \cite{Kotkin_1971}, conserves angular momentum only to leading order and fails to capture the finite rotation of the triangle. Nevertheless, the frequencies derived from this linearization do match the strongest frequencies observed in the simulation, $\sqrt{3/2}$ and $\sqrt{3}$.

Capturing the true dynamics of the system in the low-energy regime requires a description of the system in its shape subspace, as a deforming triangle instead of as three masses moving with a pairwise potential. This procedure is performed in \cite{us, Montgomery2015, Iwai2005, Littlejohn1997} for different potentials and configurations of three-body systems, by a variable change to three shape-space variables describing the shape of the triangle and one angle determining the orientation of the triangle in the plane. The choice of shape-space and orientation coordinates is a gauge choice that does not impact the results. We find that the most convenient choice for the shape variables, presented in \cite{Iwai2005}, is a Bloch sphere representation of the two relative Jacobi coordinates of the three masses,
$\boldsymbol{\rho}_1 = \sqrt{\frac{m}{2}} (\boldsymbol{r}_2 - \boldsymbol{r}_1)$
and
$\boldsymbol{\rho}_2 = \sqrt{\frac{2m}{3}}
 (\boldsymbol{r}_3 - \frac{\boldsymbol{r}_1 + \boldsymbol{r}_2}{2})$ (see Fig. \ref{fig:3masssystem}).
This variable set is denoted $\boldsymbol{w}=\left(w_{1},w_{2},w_{3}\right)$, and is given by
$w_1 = \frac{1}{2} (\boldsymbol{\rho}_1^2 - \boldsymbol{\rho}_2^2)$, related to the isometric bend mode;
$w_2 = \boldsymbol{\rho}_1 \cdot \boldsymbol{\rho}_2$, related to the asymmetric bend mode;
and
$w_3 = \boldsymbol{\rho}_1 \wedge \boldsymbol{\rho}_2$,  proportional to the oriented area of the triangle and related to the symmetric stretch mode.
 The rotation variable we use is $\theta$, describing the angle between the line connecting $m_{1}$ and $m_{2}$ and the $x$ axis. The center-of-mass coordinates decouple from the rest of the system and are set to zero. The system's invariance to rotations leads to the conservation of angular momentum $J$ and allows one to deduce the orientational dynamics, expressed via $\dot{\theta}$, from the shape-space dynamics through
\begin{equation} 
\begin{aligned}
\dot{\theta} = \frac{J}{2w} + \frac{w_{2} \dot{w_{3}} - w_{3} \dot{w_{2}}}{2w\left(w+w_{1}\right)},
\label{eq:thetadot} 
\end{aligned}
\end{equation} 
where $w=\left|\boldsymbol{w}\right|$. While $\dot\theta$ is given by the explicit relation above as a function of $\bw$ and $\dot\bw$, one can show that $\theta$ cannot be expressed as a function of $\bw$ and $\dot\bw$ alone, which in turn enables the phenomenon of deformation induced rotation. The conservation of angular momentum thus yields a non-holonomic constraint for $\theta$, and obtaining its value at a given time requires knowledge of the full dynamics of the system up to that time. As a result, $\theta$ becomes a sensitive measure for the system's dynamics and correlations.

Performing a Routh reduction of the angular momentum \cite{ArnoldDynamical1988}, we set $J=0$ to obtain the reduced shape-space Hamiltonian describing the system's zero angular momentum motion,
\begin{equation} 
\begin{aligned}
\mathcal{H}_{red}   = w\left(p_{1}^{2}+p_{2}^{2}+p_{3}^{2}\right)+
\frac{k}{2}\sum_{<ij>}\left(r_{ij}(\vec{w})-L\right)^{2},
\label{eq:reducedhamiltonian} 
\end{aligned}
\end{equation} 
where
\[
\begin{aligned}
p_{i} = \frac{\dot{w_{i}}}{2w},\qquad r_{ij}\left(\vec{w}\right)=\sqrt{2\left(w-\boldsymbol{w}\cdot\boldsymbol{b}^{ij}\right)}\,\\
\boldsymbol{b}^{13}=\left(\tfrac{1}{2},\tfrac{\sqrt{3}}{2},0\right),
\boldsymbol{b}^{12}=\left(-1,0,0\right),
\boldsymbol{b}^{23}=\left(\tfrac{1}{2},-\tfrac{\sqrt{3}}{2},0\right).
\label{eq:bij}
\end{aligned}
\]
The real space dynamics of the system can be restored by finding solutions to the reduced Hamiltonian (3) to find the shape dynamics $\boldsymbol{w}(t)$ and substituting them into Equation (2) to obtain the orientation evolution $\theta(t)$.
As proved in \cite{ArnoldDynamical1988}, every solution of the original Hamiltonian \eqref{eq:hamiltonian} with $J=0$ corresponds to a solution of the reduced Hamiltonian \eqref{eq:reducedhamiltonian}, and vice versa, therefore it suffices to study solutions of \eqref{eq:reducedhamiltonian}. 
We emphasize that the simulation results shown here are performed on the full, Cartesian system \eqref{eq:hamiltonian} while the perturbative analysis is performed on the reduced system \eqref{eq:reducedhamiltonian}.

\subsection*{A. Perturbation theory in the reduced shape space}
The reduction to shape space allows us to expand the system not about a rest position but about its equilibrium shape, the static equilateral triangle $\boldsymbol{w}_{0}=\left(0,0,\frac{mL^{2}}{2}\right)$, $\boldsymbol{p}_{0}=\left(0,0,0\right)$, thus allowing finite rotations of the triangle without breaking the small-perturbation approximation. Redefining 
\[
w^i = w_0^i+
\epsilon \alpha_i \tilde w^i,\quad 
p^i =  \epsilon {\alpha_i}^{-1}  \tilde p^i
\]
 for $\alpha_1 = \alpha_2 = \left( \frac{2 L^4 m^3}{3 k}\right)^{1/4}$, $\alpha_3 = \left( \frac{ L^4 m^3}{3 k}\right)^{1/4}$, and expanding in orders of  $\epsilon$, we obtain the Hamiltonian as a power series of the coordinates $\epsilon\boldsymbol{{\tilde w}}$ and $\epsilon\boldsymbol{{\tilde p}}$. The first non-vanishing order is the linearized Hamiltonian, quadratic in the variables and thus integrable as a simple sum of harmonic oscillators. In action-angle variables it reads,
\begin{equation} 
\begin{aligned}
\mathcal{H}_{red}  = \frac{\epsilon^{2}}{2}E_s\left(\sqrt{\frac{3}{2}}\left(I_{1} + I_{2}\right)+\sqrt{3}I_{3}\right)+\mathcal{O}\left(\epsilon^{3}\right),
\label{eq:reducedlinearizedhamiltonian} 
\end{aligned}
\end{equation} 
where $I_{j}=\frac{1}{\tau_s E_s}({\tilde w}_{j}^{2}+{\tilde p}_{j}^{2})$ are the (dimensionless) action variables serving as generalized momenta, and their conjugate coordinates are the angle coordinates denoted by $\phi_{i}$.

The nonlinearity in the system is of geometric origin rather than constitutive, and in particular is not associated with an externally  tunable expansion parameter; each of the individual springs is harmonic, and it is the geometric coupling of their strains that leads to non-linearity. As a result the nonlinear effects increase concomitantly with the strains. The largest possible strain for a given total energy is bounded and increases with the total energy.  Thus, the total energy in the system can be used to define an auxiliary expansion parameter, $\epsilon(E)$, satisfying $\epsilon(0) = 0$ and monotonically increasing with the energy.
Details of this rescaling are presented in the Supplementary Material (SM); in what follows we use $\epsilon$ as a dummy parameter in order to simplify notation, recalling that rescaling can be easily performed to yield a formal expansion for the perturbative approach in low energies.

The linearized Hamiltonian \eqref{eq:reducedlinearizedhamiltonian} describes three decoupled harmonic oscillators corresponding to the three vibrational modes of the planar triatomic molecule \cite{stomp2007colorful}: $I_{1}$ corresponds to the asymmetric stretch, $I_{2}$ to the bending mode and $I_{3}$ to the symmetric stretch (see Fig. \ref{fig:3masssystem}). We note that the 1:1 resonance between $I_{1}$ and $I_{2}$ is a result of the symmetry of the system under consideration; changing, for example, one of the masses would remove this frequency degeneracy. Substituting the solution of \eqref{eq:reducedlinearizedhamiltonian} into the equation for $\dot{\theta}_{1}$ and averaging out the fast oscillations results in the following equation for the average angular velocity \cite{Iwai2005},
\begin{equation}
\begin{aligned}
\overline{\dot{\theta}_{1}}=\epsilon^{2}\frac{3}{2\tau_{s}}\sqrt{I_{1}I_{2}}\sin\left(\phi_{2}-\phi_{1}\right).
\label{eq:linearslope}
\end{aligned}
\end{equation}
As could be inferred intuitively, overall rotation is a result of the phase difference $\phi_{2}-\phi_{1}$ between the two resonant oscillators $I_{1}$ and $I_{2}$, the asymmetric stretch and the isometric bending; the symmetric stretch oscillator $I_3$ has no rotational charge to first non-vanishing order around the equilibrium.

In a comparison to simulations, we find that this expression explains the overall angular velocities well for low enough energies, but the fit deteriorates as the energy is increased, see Fig. \ref{fig:H2phasespace}(c).  The expansion to leading order also fails to account for the beating phenomenon observed for some initial conditions (Fig. \ref{fig:typicaltraj}(b)). 
Seeking to improve the prediction for the rotation velocity as well as to account for the observed beating one can attempt canonical perturbation theory to higher orders; however, due to the 1:1 resonance, the expansion diverges already at the next non-vanishing order.

To circumvent this divergence we recast the reduced Hamiltonian in Birkhoff normal form around its static equilibrium, using the method described in \cite{ArnoldDynamical1988, Bambusi2014}. This improves the fit of the angular velocity and provides
a good description of the full dynamics observed in the regular regime, see Fig. \ref{fig:H4phasespace}.
The general procedure, presented in \cite{Bambusi2014}, is an iterative scheme of wisely chosen canonical Lie transforms that puts the system in the form of a polynomial series in action coordinates, where the series commutes with its lowest-order term. The main steps of the calculation of the normal form up to fourth order is presented in Appendix B. In the main text we present the relevant results, showing that the expansion to fourth order (presented in subsection B and C) provides an accurate description of the basic phase space structure of the full system for low energies. Furthermore, the expansion to second order suffices to break the normal mode frequency degeneracy which allows the application of Kolmogorov-Arnold-Moser (KAM) theory (subsection D).

\subsection*{B. 2$^\text{nd}$ order Birkhoff normal form}

\begin{figure*}
\includegraphics[width=\textwidth]{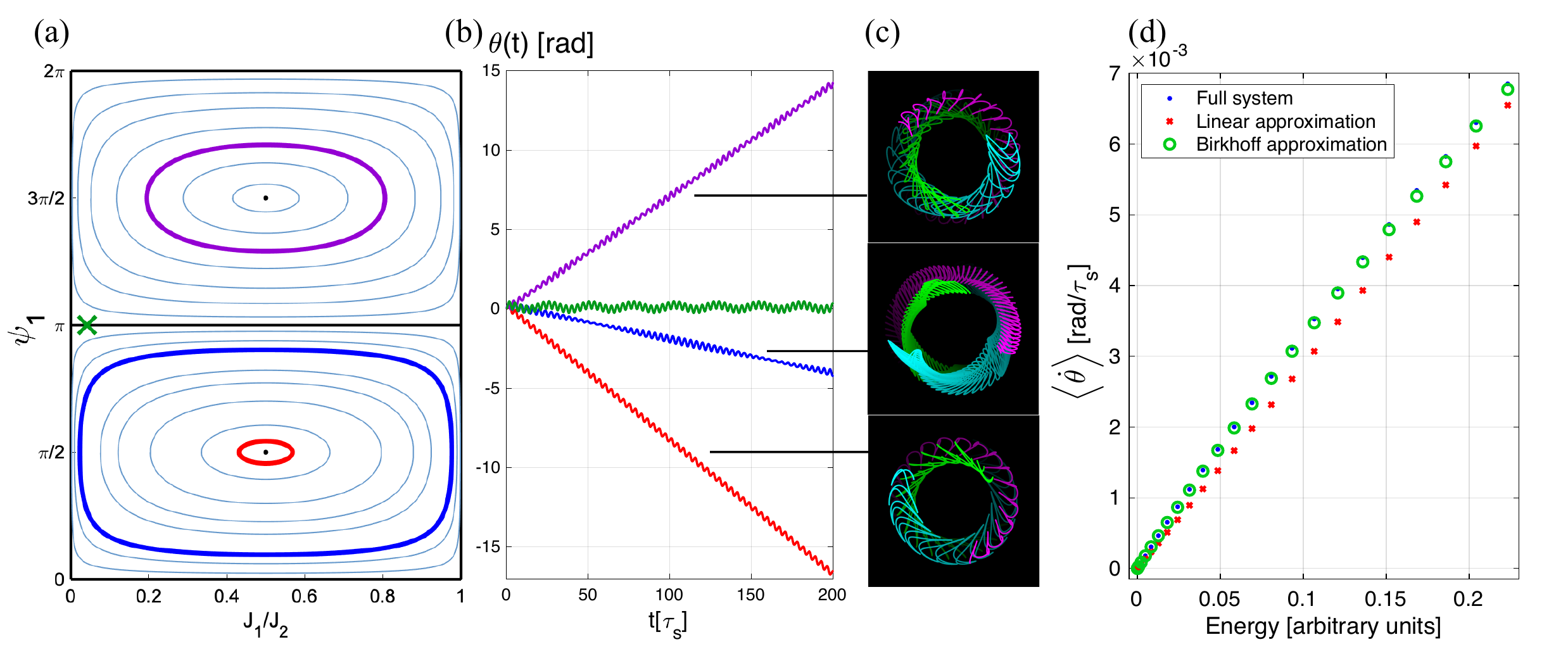}
\caption{
(a), (b) and (c) show the dynamics of the truncated system $\mathcal{H}^{(2)}$. (a) shows typical trajectories of $\mathcal{H}^{(2)}$ projected onto the $\psi_1$ vs. $J_1/J_2$ plane. (b) displays the corresponding rotations of these trajectories. (c) is a long-exposure image of their dynamics of three out of the four trajectories, with each mass colored in a different color. (d) shows a comparison of the full system's average rotation velocity as a function of the energy with the linear approximation and the much improved Birkhoff 2nd-order approximation given by $\mathcal{H}^{(2)}$.}
\label{fig:H2phasespace}
\end{figure*}

In order to recast the Hamiltonian in its Birkhoff normal form to second order, we perform a canonical change of coordinates to a new set of variables, $\left(\boldsymbol{J,\psi}\right)$, which we will use throughout the rest of this section: $J_{1}=I_{1}$, $J_{2}=I_{1}+I_{2}$, $J_{3}=I_{3}$ and their conjugate angle coordinates $\psi_{1} = \phi_1 - \phi_2$, $\psi_2 = \phi_2$, $\psi_3 = \phi_3$. These variables naturally exhibit some of the interesting behavior of the system, with $J_{3}$ as the energy contained in the area changes of the triangle, $J_{2}$ as the overall energy contained in the resonant oscillators $I_{1}$ and $I_{2}$, and $J_{1}$ as the energy contained only in $I_{1}$, satisfying  $J_{1}\leq J_{2}$. $\psi_{1}$ is the phase difference between the two resonant oscillators and describes the average rotation direction of the triangle in the plane, with $\psi_{1}\in\left(0, \pi\right)$ and $\psi_{1}\in\left(\pi,2 \pi\right)$ manifesting as counterclockwise and clockwise rotation, respectively.

In these variables, the Birkhoff normal form of the system to second order is given by
\begin{equation}
\begin{aligned}
\mathcal{H}^{\left(2\right)}=\frac{\epsilon^2}{2} E_s (H_{0}+\epsilon^{2}Z_{2})
\label{eq:H4}
\end{aligned}
\end{equation}
where
\begin{equation}
\begin{aligned}
H_{0}=&\sqrt{\frac{3}{2}}J_{2}+\sqrt{3}J_{3}, \\
Z_{2}=&\frac{1}{64} \left(52J_{1}\left(J_{2}-J_{1}\right)\sin^{2}\psi_{1}-J_{2}\left(5J_{2}+6\sqrt{2}J_{3}\right)\right).
\label{eq:birkhoff2nd}
\end{aligned}
\end{equation}
It is immediately apparent that the truncated system $\mathcal{H}^{(2)}$ is integrable, with $H_0$, $J_2$ and $J_3$ conserved quantities. Therefore, in order to visualize the dynamics of the truncated system, it suffices to consider the two-dimensional phase plane $\frac{J_1}{J_2}$ - $\psi_1$
given values $J_{3}\geq 0$ and $J_{2}>0$\footnote{When $J_{2}=0$, $J_{1}$ vanishes as well, corresponding to equilateral triangles. This is an integrable family of special symmetries, with no discernible impact on the observable phase space when $J_2>0$.}, as seen in Fig. \ref{fig:H2phasespace}. This phase plane has the structure of a finite cylinder, with $0 \leq {J_1}/{J_2} \leq 1 $ and $\psi_1 \in [0, 2 \pi]$ an angle variable, and can be unfolded onto the plane (see Fig. \ref{fig:H2phasespace}). The system has nullclines at $J_{1}/J_2=0$ and $J_{1}/J_2=1$, fixed lines at $\psi_{1}=k \pi$ for $k\in\mathbb{N}$ and two distinct elliptic fixed points at $\left(J_{1}/J_2,\psi_{1}\right) = \left(1/2,\pi/2\right)$ and $\left(J_{1}/J_2,\psi_{1}\right) = \left(1/2,3\pi/2\right)$. Solutions of $\mathcal{H}^{(2)}$ perform closed orbits around the distinct fixed points and cannot cross the rectangles drawn by the nullclines and the fixed lines. These trajectories display a beating phenomenon where energy periodically transfers between the resonant oscillators, with more energy contained in $I_1$ ($I_2$) when $J_1/J_2<1/2$ ($J_1/J_2>1/2$).

The integrable dynamics of the truncated system lie in the shape space of the system, and can be pulled back to obtain the corresponding rotation of the triangle in the plane. As seen in Fig. \ref{fig:H2phasespace}, rotation around the fixed point $\psi_1 = \pi/2$ manifests as a negative angular velocity, while rotation around the other fixed point $\psi_1 = 3\pi/2$ as positive angular velocity. Trajectories that are close to the fixed points have a smaller beating frequency and a larger absolute angular velocity than trajectories that pass closer to the rectangles' borders, in which the beating is very apparent, for the same energy. Despite the beating, since the motion around the fixed points is periodic, the averaged rotational velocity of a given trajectory is constant in all cases.

\subsection*{C. Birkhoff normal form to higher orders}
\begin{figure*}
\includegraphics[scale=0.7]{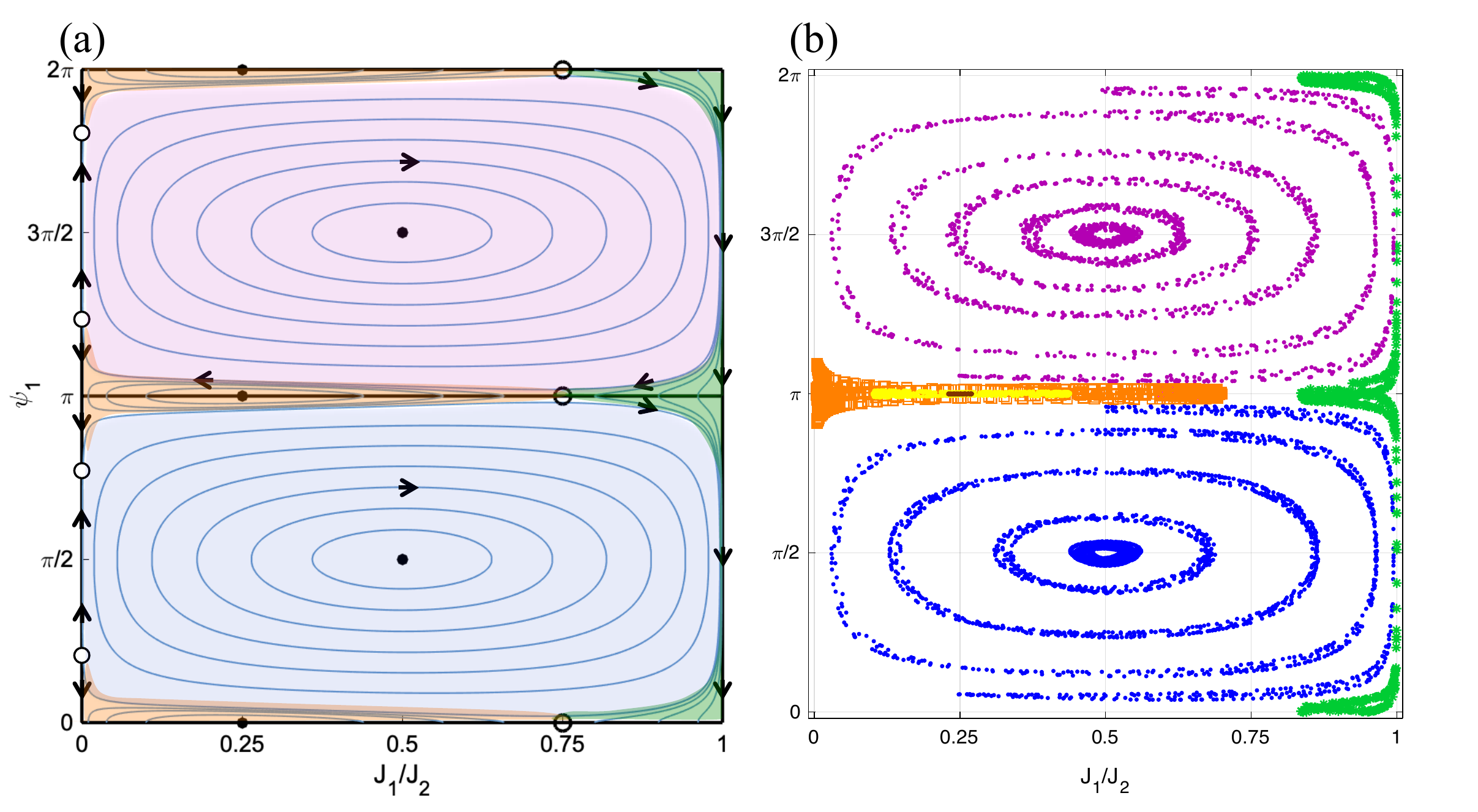}
\caption{
(a) Phase space of $\mathcal{H}^{(4)}$ compared with (b) Poincar\'e sections of the full system \eqref{eq:hamiltonian} at a low energy $E=0.0075$.
(a) is divided into four dynamical regions: the purple region, in which trajectories encircle the fixed point $J_1 = J_2/2, \psi_1 = 3\pi/2$, describing clockwise rotation of the triangle; the blue region of trajectories encircling $J_1 = J_2/2, \psi_1 = \pi/2$ describing counterclockwise rotation; the orange region of trajectories encircling $J_1 = J_2/4, \psi_1 = 0$; and the green region of trajectories migrating along the nullcline $J_1 = J_2$. (b) shows that for low enough energies, the full system follows the regular structure of $\mathcal{H}^{(4)}$ to a high level of accuracy, with six typical trajectories shown in the purple region, five in the blue region, three in the orange region and one in the green region.. }
\label{fig:H4phasespace}
\end{figure*}
Of course, a priori there is no guarantee that the normal form should describe or even approximate the dynamics of the full system that we observe in simulations. The normal form series does not necessarily converge, and the theory is guaranteed to hold only for some neighborhood of the static equilibrium configuration. We can check compatibility by comparing the simulated dynamics of the full system \eqref{eq:hamiltonian} with the predicted normal-form dynamics of $\mathcal{H}^{\left(2\right)}$, projecting the trajectories onto the unfolded $J_1/J_2$ - $\psi_1$ plane. As can be seen in Fig. \ref{fig:H4phasespace}b, the two elliptic fixed points of $\mathcal{H}^{\left(2\right)}$ are clearly visible in the full system. 
Trajectories of the full system that initialize close enough to the fixed points shadow the truncated system's trajectories, with similar beating frequencies and a similar overall angular velocity of the rotating triangle (Fig. \ref{fig:H2phasespace}c). However, despite this high compatibility, some prominent elements of the dynamics are not captured by $\mathcal{H}^{\left(2\right)}$. The lines $\psi_1 = k \pi$ for integer $k$, are not fixed for the full system; rather, simulations of the full system show an elliptic fixed point at 
 $\left(J_{1}, \psi_{1}\right) = \left(J_{2}/4, \pi\right)$, around which beating trajectories with zero overall angular velocity are distinctly apparent in the simulations. 

In order to capture these features, we now come to consider the Birkhoff normal form to the next non-vanishing order, given by
\begin{equation}
\begin{aligned}
\mathcal{H}^{\left(4\right)}=\frac{\epsilon^2}{2} E_s (H_{0}+\epsilon^{2}Z_{2} + \epsilon^{4}Z_{4}),
\label{eq:H4}
\end{aligned}
\end{equation}
where
\begin{equation}
\begin{aligned}
Z_{4}=&a_{0}+J_{1}\left(a_{1}+b_{1}\cos2\psi_{1}\right)+J_{1}^{2}\left(a_{2}+b_{2}\cos2\psi_{1}\right) \\
&+J_{1}^{3}\left(a_{3}+b_{3}\cos2\psi_{1}\right).
\label{eq:birkhoff}
\end{aligned}
\end{equation}
The coefficients $\left\{ a_{i},b_{i}\right\} _{i=0}^{3}$ are functions of $J_{2}$ and $J_{3}$, presented in full in Appendix 3 along with the full calculation. $\mathcal{H}^{\left(4\right)}$ still conserves $J_{2}$ and $J_{3}$, retaining the integrability of the normal form.

The distinct elliptic fixed points of  $\mathcal{H}^{\left(2\right)}$ are also fixed points in the phase space of $\mathcal{H}^{\left(4\right)}$. However, the $\psi=k \pi , \; k \in \mathbb{Z}$ lines lose their stability; instead, two new fixed points emerge on each line, an elliptic fixed point at $\left(J_{1},\psi_{1}\right) = \left(\frac{J_{2}}{4},k \pi\right)$ and a hyperbolic fixed point at $\left(J_{1},\psi_{1}\right) = \left(\frac{3J_{2}}{4},k \pi \right)$. Also, four hyperbolic fixed points appear on the nullcline of $J_{1}=0$; see Appendix C for the full fixed point analysis. The fixed points separate the trajectories into two classes of trajectories, those encircling the elliptic fixed points and those migrating along the nullclines of $J_{1} = 0$ and $J_{1} = J_{2}$. 

Pulling back from shape space to the space of triangles in the plane, the rotation resulting from the dynamics of $\mathcal{H}^{\left(4\right)}$ are similar to that of $\mathcal{H}^{\left(2\right)}$ close enough to the fixed points shared by the systems. However, the dynamics are different around the new phase space features: trajectories going around $\left( J_{2}/4 , k\pi \right)$ periodically rotate in both directions in real space, with an overall vanishing averaged rotation rate. The trajectories following the nullclines also do not perform overall rotation. The hetroclinic trajectories between the hyperbolic fixed points define the boundary between the different dynamical regions.
As can be seen in Fig. \ref{fig:H4phasespace}, these dynamics are indeed compatible with the full system to a high degree. Poincar\'e sections of the full dynamics projected onto the $\left(J_{1}/J_2,\psi_{1}\right)$ plane reveal exactly the same fixed points as calculated from $\mathcal{H}^{\left(4\right)}$. 

\subsection*{D. Lifting the frequency degeneracy}

The harmonic three mass system shows a strong persistence of regular solutions for a large range of energies. This suggests the applicability of the KAM theorem, which guarantees persistence of most quasi-periodic orbits in almost-integrable systems if the integrable part satisfies some non-degeneracy frequency condition. Unfortunately, the harmonic expansion (to lowest order), $H_{0}$, shows a one-to-one resonance, and thus cannot serve as the base of a KAM expansion. 
 Therefore, to express our system as an almost-integrable system, we write the full Hamiltonian as  $\mathcal{H} = \mathcal{H}^{(2)} +( \mathcal{H} - \mathcal{H}^{(2)})$. The Birkhoff expansion, detailed in the SM, shows that the remainder in the parentheses is a power series, $\mathcal{H} - \mathcal{H}^{(2)} = \sum_{n=2}^\infty P^{(n)}(\boldsymbol{J}, \boldsymbol{\psi})$ where $P^{(n)}$ is a monomial of order $n$ in the action coordinates. The auxiliary expansion parameter $\epsilon(E)$, a monotonically decreasing function of the maximal possible stretch given the energy $E$ (see SM), can then be used to rescale the action coordinates, so that $0\leq J_i\leq 1$ for $i=1,2,3$. Thus, the full Hamiltonian is rewritten as a power series in $\epsilon(E)$ multiplying terms of order $1$.
Hence, so long as the entire remainder is small, we can treat our system as an almost-integrable system, considering  $( \mathcal{H} - \mathcal{H}^{(2)})$ as the perturbation to the integrable and non-degenerate $ \mathcal{H}^{(2)}$ .

The KAM theorem does not provide a realistic bound on what consists a small enough perturbation for theorem to be applicable.
Nevertheless, we can check the frequency conditions required for the theorem by performing a canonical change of variables to action-angle variables and calculating the frequencies:x
\begin{equation}
\begin{aligned}
\mathcal{H}^{\left(2\right)}=&\frac{\epsilon^2}{2} E_s \left( \sqrt{\frac{3}{2}}J_{2}+\sqrt{3}J_{3}\right.\\
&\left.-\frac{\epsilon^{2}}{64}\left(5 J_{2}^{2}-6\sqrt{2} J_{2}J_{3}+13  K_{1}^{2}\right)\right);
\label{H2AA}
\end{aligned}
\end{equation}

\begin{equation}
\begin{aligned}
\omega_{1}	=&\,\epsilon^{2}\frac{13}{32} K_{1} ,\\
\omega_{2}	=&\sqrt{\frac{3}{2}}
-
\frac{\epsilon^{2}}{32}
\left(
\left(5-3\sqrt{2}\right) J_{2}+
13 K_{1}\right), \\
\omega_{3}	=&\sqrt{3}+\epsilon^{2}\frac{3\sqrt{2}}{32}J_{2};
\label{eq:birkhoffFrequencies}
\end{aligned}
\end{equation}
where
$K_{1} = 2\sqrt{J_{1}\left(J_{2}-J_{1}\right)}\sin\psi_{1}$
 is the new conserved quantity emerging from the integrable system. Note that it is proportional to the linear slope prediction Eq. \eqref{eq:linearslope}; indeed, the sign of $K_1$ indicates the overall direction of rotation of the triangle, see Fig. \ref{fig:Levy}(c).

The new frequencies associated with the angle coordinates have corrections of order $\epsilon^2$ which depend on the action coordinates, thus removing the degeneracy of the linearized system. 
It is easy to check that both the non-degeneracy and the isoenergetic non-degeneracy conditions stated in the KAM theorem are satisfied for small enough values of $J_{2}$, $J_{3}$. Under these conditions, the KAM theorem assures that most integrable tori persist under small perturbations to the Hamiltonian for any energy value that is small enough. 
Taking into account $\mathcal{H}^{(4)}$ as the integrable part would add corrections of order $\epsilon^4$, retaining this degeneracy lifting.

The loss of integrability is expected to manifest first around resonant tori, overtaking most of the phase space gradually as the perturbation grows. This picture is compatible with our numerical experiments and provides a possible explanation for the good fit between the truncated and the full system's dynamics for low enough energies. For short times, this shadowing of the trajectories of the truncated integrable Hamiltonian by the full Hamiltonian trajectories remains as the energy is further increased, as we show next.

\subsection*{E. The L\'evy Walk regime }

\begin{figure}
\includegraphics[width=0.5\textwidth]{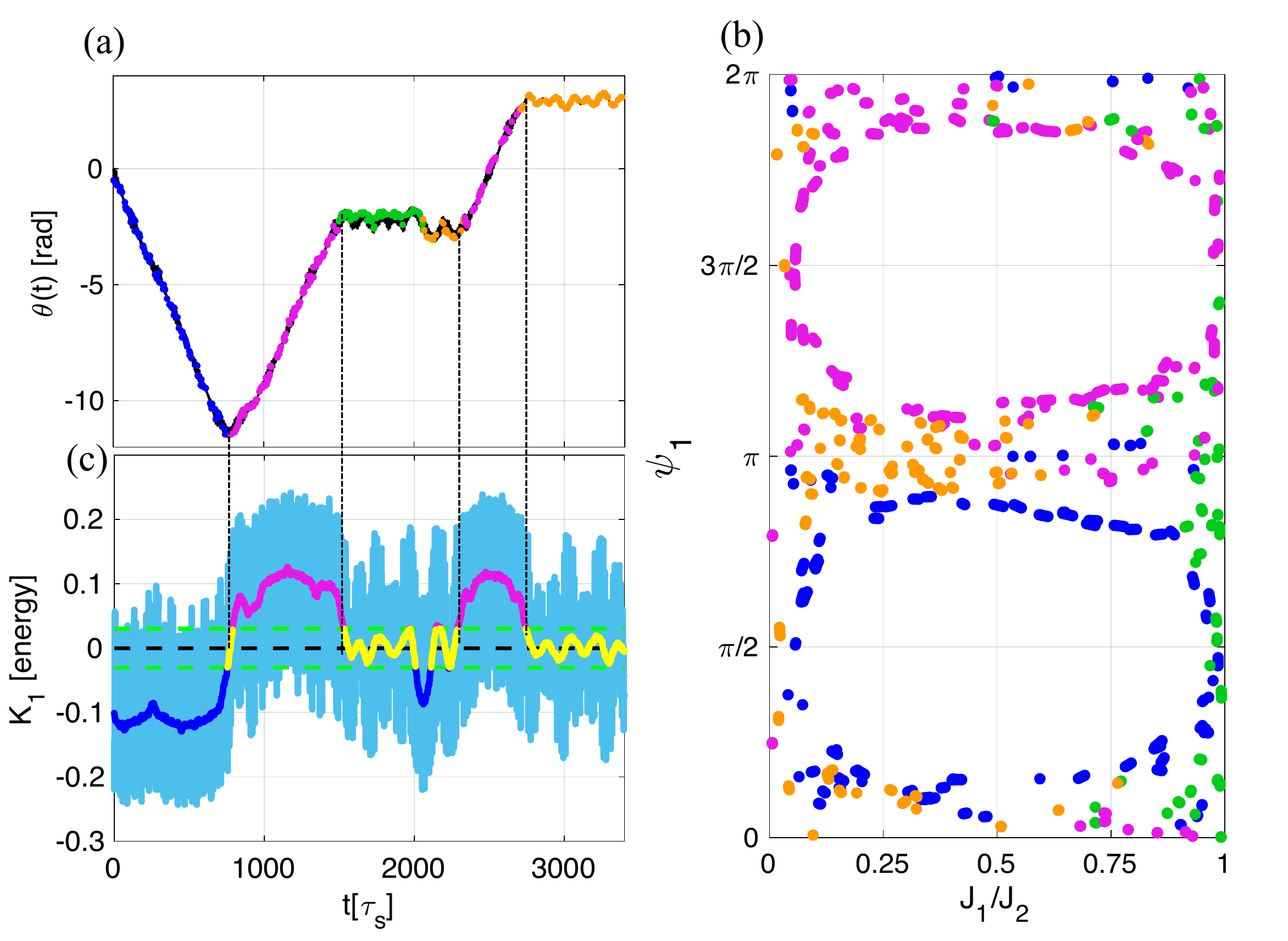}
\caption{(a) A typical orientational trajectory at energy $E = 0.381$. (b) Projection of the phase space dynamics onto a Poincar\'e section of the $(J_1/J_2, \psi_1)$ plane. The power-law statistics observed in this energy regime correspond to a sticking of the irregular trajectories close to the approximated integrable system's fixed points for long times.
(c) The action variable $K_1$ emerging from $\mathcal{H}^2$ is shown in light blue, with its moving average $\langle K_1 \rangle$ on top, averaged over 100 time units. $\langle K_1\rangle<-0.03 $ is colored in blue and corresponds to a descending rotation angle; $\langle K_1\rangle > 0.03$ is in magenta and corresponds to an increasing rotation angle; and $-0.03 < \langle K_1\rangle < 0.03$ is in yellow and corresponds to zero averaged rotation. $K_1$ does not differentiate between trajectories that encircle the $J_1/J_2 = 0.25$ (orange in (a,b)) and trajectories that travel along the $J_1/J_2 = 1$ border (green in (a,b)), but may identify a transition between them. 
It is obvious from (c) that $K_1$ is not a conserved quantity of the full Hamiltonian, nor is it a constant along seemingly ballistic bouts. Nonetheless, its moving average is indicative of the different regimes.
}
\label{fig:Levy}
\end{figure}

At energies in the range $E_{s}/15\lesssim E\lesssim E_{s}/9$ most trajectories are no longer regular: chaotic dynamics characterized by a positive Lyapunov exponent inhibit most of phase space. At the early stages of this regime, the corresponding rotational dynamics resemble the stochastic L\'evy-walk model \cite{us}, with bouts of constant angular velocity interrupted by abrupt orientation reversal events.
In Fig. \ref{fig:Levy} we plot the angular dynamics in this regime, alongside a  projection of phase space onto the $\left(J_1/J_2,\psi_1\right)$ plane, where the system is shown to follow the integrable structure described by $\mathcal{H}^{\left(4\right)}$.
The projection indicates that the observed trajectories migrate between the different fixed points of $\mathcal{H}^{\left(4\right)}$, sticking to oscillatory trajectories around each of the stable fixed points for long times. We further observe that the transitions between the distinct neighborhoods of the fixed points occurs near the saddle points located at $J_{1}/J_{2}=3/4, \; \psi=\pi k$ for $k \in \mathbb{Z}$, and the transition times obey a power law distribution.

At the lowest energies in which we observe L\'evy-walks each bout between orientation reversal events bears great resemblance to the corresponding regular trajectory around the same fixed point, to the extent that it is difficult to differentiate between regular and L\'evy-walk trajectories just by examining them for short times in between transitions.
As the energy is increased  the transitions between the neighborhoods of the fixed points become more frequent and occur over an increasingly wider region. As the energy approaches $E_{s}/9$ from below it seems that there is no longer any barrier separating the 
basins of the distinct fixed points, and the bouts gradually lose their coherence and similarity to the regular solutions. Nonetheless, the squared angular mean displacement still obeys fractional statistics \cite{us}.

In low-dimensional systems, $d\leq 2$, the emergence of power-law statistics is well-understood, attributed to the breakdown of Kolmogorov-Arnold-Moser (KAM) tori creating partial transport barriers which can be crossed by chaotic trajectories at a slow rate in a phenomenon commonly referred to as sticking or trapping \cite{lange2016mechanism, mackay1984transport, meiss1986markov, cristadoro2008universality, AFM14, alus2017universal}.
However, despite the robustness of this phenomenon \cite{latora1999superdiffusion, cagnetta2015strong, Shlesinger1987, Zaburdaev2015, zaslavsky2002chaos},  a general framework for the origin of power-law statistics in high-dimensional mixed Hamiltonian systems continues to elude current understanding \cite{lange2016mechanism, danieli2017intermittent, das2019power, meiss2015thirty}. In the three-body harmonic system, the coherent bouts creating the power-law statistics strongly resemble their lower-energy regular counterparts, indicating a partial trapping of chaotic trajectories around regular islands for finite times. A quantitative analysis of the dynamical mechanism behind this phenomenon would require a deep understanding of the regular behavior of the non-linear system.

In mixed Hamiltonian systems, fractional statistics are ubiquitous \cite{latora1999superdiffusion, cagnetta2015strong, Shlesinger1987, Zaburdaev2015, zaslavsky2002chaos}. In systems with two degrees of freedom, where regular tori create barriers in phase space, the origin of these anomalous statistics is well understood. Generally, as the KAM tori break up, they leave in their wake a hierarchical structure of smaller tori that create partial barriers of transport. Chaotic trajectories can cross these barriers, but this typically takes a long time, resulting in the fractional statistics \cite{meiss1986markov, meiss2015thirty}. However in higher dimensional systems, the phase space mechanism creating and controlling the observed fractional statistics is not yet fully understood \cite{lange2016mechanism}. Although the KAM tori break up in a similar manner, they no longer separate phase space into impenetrable regimes. Thus chaotic trajectories can theoretically get as close as they like to the surviving tori. In these systems, for any perturbation strength, the phase space is connected by a web of resonant channels known as the Arnold web surrounding the sufficiently non-resonant KAM tori. Action variables can drift along these channels in a process known as Arnold diffusion and thus transition from the neighborhood of one surviving torus to another. In our system, the great resemblance of the low-energy L\'evy-walk trajectories to regular solutions and the narrow channel of transfer are reminiscent of the Arnold diffusion phenomenon.  On the other hand, as the energy rises and the transition region grows, the $\mathcal{H}^4$ phase space structure loses its coherence and the power-law statistics seem to originate from a partial trapping of trajectories around the regular fixed points.
A combination of the two phenomenon could explain the surprising phenomenon of a gradual, seemingly continuous decrease of the anomalous exponent from the ballistic to the random walk regime as the energy grows, as observed in \cite{us}, as opposed to the single anomalous exponent found in \cite{alus2017universal, shepelyansky2010poincare}. This work provides the backbone that would be required for a systematic study of these concepts, by identifying the underlying almost-integrable approximation controlling the dynamics in the transition of the full system from regular behavior to chaos. These allow a calculation of the KAM tori and the surrounding Arnold web. A quantitative study of these ideas is left to future work.

\section{High Energy regime}

As the energy is increased beyond $E_s/15$, the regular structure gradually disappears. In the range $E_{s}/9\lesssim E\lesssim2E_{s}$, almost all trajectories are observed to cover the entirety of phase space, and the statistics resemble an uncorrelated random walk \cite{us}.

However, at $E\approx 2E_{s}$ a single frequency begins to dominate the dynamics and the system appears to approach regularity again. This apparent regularity may be easily explained by observing that for extremely high energies, the rest length is effectively forgotten. Thus we may expect the system to resemble the integrable harmonic three-mass system with zero rest lengths \cite{CS93} at high enough energies.
In this regime the reduced system \eqref{eq:reducedhamiltonian} is less instructive, since its normal modes do not coincide with those of the zero rest length harmonic three-mass problem.
Therefore, we compare the observed dynamics with the dynamics of the full Cartesian Hamiltonian \eqref{eq:hamiltonian} with zero rest lengths:
\begin{equation}
\begin{aligned}
\mathcal{H}=\sum_{i=1}^{3}\frac{\p_{i}^{2}}{2m}+\sum_{<ij>}\frac{k}{2}r_{ij}^{2},
\label{eq:Hhigh}
\end{aligned}
\end{equation}
which displays the twice-degenerate linear frequency $\sqrt{3}$ in units of $1/\tau_s$.

\begin{figure}
\includegraphics[width=0.45\textwidth]{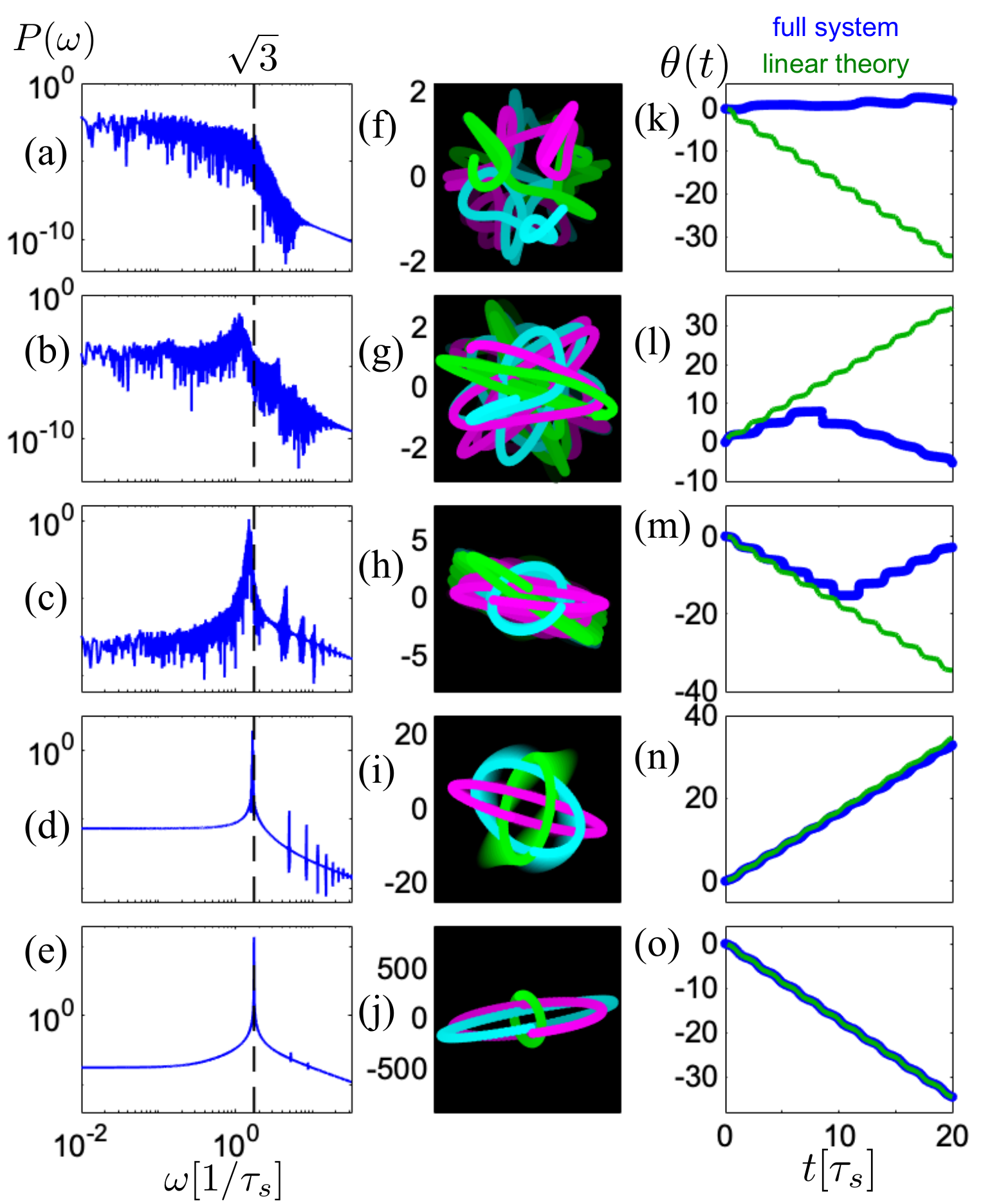}
\caption{(a-e) Power spectrum of the $x$ coordinate of $m_1$ for various typical trajectories with rising energies values from $E\sim 1$ to $E\sim 10^6$. (f-j) Long-exposure images of the corresponding trajectories. (k-o) Orientation as a function of time of the full system (blue) and the linear prediction obtained by solving the zero-rest length system with the same initial conditions (green). As the energy gets larger, the high-energy limit is a better approximation to the full system dynamics. The energy values of the different plots are (a,f,k) $E = 1.47$, (b,g,l) $E = 9.51$, (c,h,m) $E = 95.42$, (d,i,n) $E = 1354$, (e,j,o) $E = 2.11\times 10^6$.}
\label{fig:highenergy}
\end{figure}

As shown in Fig. \ref{fig:highenergy}o, the solutions of \eqref{eq:Hhigh} are in excellent agreement with the simulation results for high enough energies.
Similarly to the low energy regime, the regular solution of the high energy regime displays a constant average angular slope. However, unlike the solution for low energies, this slope is comprised of steps: discrete angular increment events in between which the angle remains approximately constant. This feature may be explained by observing that for high energies, at every oscillation the three-mass triangle undergoes two orientation reversal events. 
Both orientation reversal events occurring in a single oscillation increase (or decrease, depending on initial conditions) $\theta(t)$ by $\pi$. Hence the expected averaged slope has a constant value of $\sqrt{3}\approx 1.73$, as observed.

The step structure is preserved at moderately high energies, where the rest lengths are not completely negligible, and is observed even at energies that are very close to the energy scale, $E \gtrsim 2 E_s$, see Fig.  \ref{fig:highenergy} (l,m,n). For these moderate energies, the dominant frequency drifts away from $\sqrt{3}$ and the spectrum fills up (Fig.  \ref{fig:highenergy} b,c,d), and for low enough energies $\theta(t)$ changes its average rotation direction in a manner resembling the L\'evy walk region. 
A statistical and perturbative analysis of the Hamiltonian in this energy regime is left to future studies, but it is clear that a similar approach to the perturbative techniques used in the low energy regime may be useful in analyzing the approach of the system to high energies where the system behaves like its linear approximation.

\section{Summary and Discussion}

Despite the apparent simplicity of the harmonic three-mass system with finite rest lengths, the system displays a rich variety of dynamics, controlled mainly by the system's overall energy. For very low energies the system displays constant angular velocity rotation with zero angular momentum, while for moderate energies chaos ensues, indicated by a positive Lyapunov exponent, and an orientational random walk is measured. Gradually increasing the energy from very low to moderate values reveals statistics of an orientational L\'evy walk, in which the exponent $\alpha$ continuously varies with the energy, interpolating between the values $\alpha=2$ (ballistic rotation) and $\alpha=1$ (random walk). Further increasing the energy beyond the random walk region, the system gradually  ``forgets'' its finite rest lengths and the systems trajectories regain regularity.

In the chaotic regimes, while the observed trajectories share many characteristics with regular solutions of integrable approximations of the Hamiltonian, no explicit solutions are available. Thus identifying the fixed points of the system's Hamiltonian, the structure of the regular solutions and the geometry of the phase space for the integrable approximations of the  Hamiltonian is key to understanding the exotic phenomena that the full system displays.
Furthermore, observing the rich phenomena that the harmonic three mass system displays requires to simulate the underlying chaotic Hamiltonian system to very long times ($\sim 10^{7}$ Lyapunov times). Such a task is not commonly carried out, primarily because of the difficulty in interpreting the result and identifying the real system it describes \cite{Yao2010}. Understanding analytically the building blocks from which the different parts of the full trajectory is composed not only allows to understand the origin of the observed L\'evy walks, but also serves to cast meaning to the observed trajectories as typical members in a collection of statistically similar trajectories that cover the chaotic component of phase space.
To these ends, in this work we mapped and characterized the fixed points and regular solutions of both the low and high energy integrable approximations of the Hamiltonian of the harmonic three mass system.

Very low energy trajectories display a constant angular velocity rotation with zero angular momentum that is well captured by reducing the system to its intrinsic (shape) space and 
linearizing the system about its equilibrium shape. As the energy is increased non-linear effects including lifting of the frequency degeneracy and beating dramatically change the value of the constant angular velocity. Capturing these variations required a perturbative approach.
Due to the 1:1 resonance in the system's linearization, canonical perturbation theory diverges, therefore we utilized the Birkhoff normal form expansion. Expanding the Hamiltonian in Birkhoff normal forms to 4th order yielded a very good agreement with the observed average angular velocity, and was shown to accurately predict the phase space structure observed for the full system. The action angle variables inherited from these approximations capture the behavior of the system near the onset of chaos. For higher energy, when L\'evy walks become more pronounced these action variables are no longer constant even along seemingly ballistic bouts. Nonetheless, the predicted phase space structure still underlies the full dynamics, and the L\'evy walks can be decomposed to bouts that dwell in the vicinity of regular trajectories rotating clockwise and anticlockwise at a constant pace, and power-law distributed transition between these trajectories. 

As the system approaches regular random walk with $\alpha=1$ the structure of phase space predicted from the integrable approximations ceases to describe the system. For a narrow strip of energies in the random walk region the system appears lacking an underlying structure. However, as the energy of the system is further increased a new structure emerges. When the typical mass separation significantly exceeds the rest length, $L\ll  \langle |r_{ij}|\rangle$, the rest lengths are effectively lost, and for $10^{4} E_{s} \lesssim E$ the observed trajectories seem regular again, and can be explained considering the system with vanishing rest lengths. This new structure begins to be apparent already at energies $E_{s} < E$ as a single frequency starts to dominate the power spectrum of the system, while still in the chaotic regime.

For energies in the intermediate range $E_s/15\lesssim E\lesssim E_s/9$, anomalous power-law statistics of the system's variables are measured \cite{us}. 
The phase-space mechanism behind power-law correlations and corresponding anomalous diffusion of measurables in systems with a high phase space dimension is not well-understood, and may be attributed to Arnold diffusion, stickiness or some combination of the two \cite{lange2016mechanism}. 
The quantitative understanding of the regular structure achieved in this work is crucial in order to understand and quantify the anomalous region, and to differentiate between the mechanisms responsible. In our system we observe that at the onset of the region, trajectories spend long times circling one of the low-energy Birkhoff expansion fixed points, resembling their corresponding regular trajectories, before transitioning to a different fixed point through a narrow transfer channel around a hyperbolic fixed point. This scenario is reminiscent of an Arnold diffusion mechanism. As the energy rises, the KAM islands shrink and this description gradually loses its coherence,
resulting in an anomalous exponent that appears to interpolate smoothly between ballistic and random-walk values \cite{us}. 
While we presently cannot prove so, we believe both Arnold diffusion and sticky dynamics dominate the system's behavior for different energies, partially explaining the smooth interpolation between the ballistic and regular diffusive regimes.   

*************

\newpage

\section*{Appendices}

\subsection{Obtaining a small parameter $\epsilon(E)$}
\label{App:epsilon}

In formulating the system using perturbation theory, $\epsilon$ was an auxiliary variable that only served as a dummy parameter to ease the expansion.
However, using the geometrical constraints of the system we can provide an estimate for $\varepsilon\left(E\right)$, and use it to rescale the parameters so that they remain bounded. This is a special feature of the spring-mass system, as opposed to some other chaotic systems such as the three-body gravitational problem: since the full Hamiltonian is positive-definite in its parameters, the overall energy of the system limits the kinetic energy that the masses can gain, and because of the geometry the masses cannot drift farther away from each other than a certain radius. This allows us to place bounds on the action variables given an energy value $E$,
\[
\begin{aligned}
&0
\leq
I_{1}
\leq
\frac{4 \sqrt{\frac{2}{3}} E}{3 k \left(L-\sqrt{2} \sqrt{\frac{E}{k}}\right)^2},
\\
&0
\leq
I_{2}
\leq
\frac{4 \sqrt{\frac{2}{3}} E}{3 k \left(L-\sqrt{2} \sqrt{\frac{E}{k}}\right)^2},
\\
&0
\leq
I_{3}
\leq
\frac{4 E }{3 \sqrt{3} k \left(L-\sqrt{2 \frac{E}{k}}\right)^2}
\label{eq:ilimits}
\end{aligned}
\]
We thus define $\epsilon(E)$ to be the larger of the three:
\begin{equation} 
\begin{aligned}
\varepsilon\left(E\right)^{2} = 
\frac{4 \sqrt{\frac{2}{3}} E}{3 k \left(L-\sqrt{2} \sqrt{\frac{E}{k}}\right)^2},
\label{eq:epsilonofE} 
\end{aligned}
\end{equation} 
defined so that at a given energy $E$ the action variables $I_{j}$ cannot surpass $\varepsilon\left(E\right)^{2}$. 
An estimate of the energy at which perturbation theory is expected to break down is given by comparing $\varepsilon\left(E\right)$ to $1$, occurring at $E\approx0.66$. Indeed as numeric show, this value is close to the energy at which we see an onset of chaos.

Further, by rescaling the action parameters $I_{j} = \varepsilon\left(E\right)^{2}\tilde{I}_{j}$, we know that their range is always $0\leq\tilde{I}_{j}\leq1$, and $\varepsilon\left(E\right)$ is a monotonically increasing function of $E$, satisfying $\epsilon(E=0) = 0$. Therefore, for small enough energies the bulk of the energy is contained in low orders of the $\varepsilon$ expansion, constraining the remainder and providing further justification of the applicability of perturbation theory techniques to analyze the system as nearly-integrable.

\subsection{Birkhoff Normal Form to 6th Order}\label{App:red}

Given an m-dimensional Hamiltonian $\mathcal{H}$ with an elliptic fixed point at the origin, consider the linearized Hamiltonian about its fixed point, $H_0 = \sum_{i=1}^m \omega_i \frac{p_i^2 + q_i^2}{2}$. Then the Birkhoff normal form theorem states that 
for any positive integer $N\geq0$ there exists a neighborhood 
$\mathcal{U}_{N}\subset\mathbb{R}^{2n} $
of the origin and a canonical transformation 
$\mathcal{T}_{N}:\mathcal{U}_{N}\rightarrow\mathbb{R}^{2n} $
that brings the full system to its Birkhoff normal form up to order N:
\begin{equation}
H^{\left(N\right)}:=H\circ\mathcal{T}_{N}=H_{0}+Z^{\left(N\right)}+\mathcal{R}^{\left(N\right)}
\end{equation}
where
$Z^{\left(N\right)}$ is a polynomial of degree N+2 that Poisson commutes with its leading order expansion about the fixed point, $H_{0}$, i.e.
$\left\{ H_{0},Z^{\left(N\right)}\right\} \equiv0$ , and 
$\mathcal{R}^{\left(N\right)}$ is small, i.e.
$\left|\mathcal{R}^{\left(N\right)}\left(x\right)\right|\leq C_{N}\left|x\right|^{N+3}\,,\,\,\forall x\in\mathcal{U}_{N}$.

A proof of this theorem is given in \cite{Bambusi2014}. It is a constructive proof with a general recipe for obtaining the Birkhoff normal form up to any desired order $N\in\mathbb{N}$, given a Hamiltonian with an elliptic fixed point at the origin. Here we present the main steps of the construction for our system.

The recipe is based on a series of Lie coordinate transforms chosen such that the polynomial correction $Z^{(N)}$ Poisson commutes with $H_0$.
A Lie transform of coordinates is a canonical change of variables induced by some generating function $\chi$.
Assume we have a polynomial $g(p,q)$ of order $n+2$, and a Lie transform generator $\chi(p,q)$, which is a polynomial of order $m$. 
Consider $\phi_\chi^t = (p(t), q(t))$, the propogation of the variables $p$ and $q$ according to a Hamiltonian given by $\chi$. We seek to express the original polynomial $g(p,q)$ estimated at the propogated coordinates: $g(p(t), q(t)) \equiv g\circ \phi_\chi^t$.
Setting $t=1$, the new polynomial can be written as a power series in the order of the polynomials,
\begin{equation}
g\circ\phi_\chi^1=\sum_{k\geq0}g_{k}
\end{equation}
where
\begin{equation}
g_{0}:=g\,\,,\,\,g_{k}=\frac{1}{k}\left\{ \chi,g_{k-1}\right\} \,,\,k\geq1
\end{equation}
and the order of the polynomial $g_k$ is $n+k m$.

Consider now a polynomial Hamiltonian expanded in powers of the coordinates and momenta about its elliptic fixed point, $\mathcal{H} = \epsilon^2 H_0(p,q) + \sum_{n=1}^\infty \epsilon^{n+2} P_n(p,q)$, where $P_n(p,q)$ is a sum of monomials of order $n+2$, of the form $q^L p^{n+2-L}$. This Hamiltonian is already in Birkhoff normal form to zeroth order.
For any first-order polynomial $\chi_1$, the corresponding Lie transform of $\mathcal{H}$ leads to the ordered form:
\begin{equation}
\begin{aligned}
\mathcal{H} \circ \phi_{\chi_1} = &\epsilon^2 H_0 +\\ 
&\epsilon^3 (P_1 + \{ {\chi_1}, H_0 \}) + \\
&\epsilon^4 (P_2 + \{ {\chi_1}, P_1 \} + \{ {\chi_1}, \{ {\chi_1}, H_0 \} \}) + \mathcal{O}(\epsilon^5)
\label{eq:Birkhoff1}
\end{aligned}
\end{equation}
As $P_1 +   \{ {\chi_1}, H_0 \} $ is a polynomial of order $3$, choosing $\chi_1$ such that this term commutes with $H_0$ will bring Eq. \eqref{eq:Birkhoff1} to its Birkhoff normal form up to 1st order.

In general, obtaining an $n$'th degree Birkhoff normal form is done iteratively. Consider a
Hamiltonian given in its Birkhoff normal form up to order $n-1$, i.e. $\mathcal{H} \circ \mathcal{T}_{n-1} = \epsilon^2 H_0 + Z^{(n-1)} + \mathcal{R}^{(n-1)}$: $Z^{(n-1)}$ is a polynomial of order $n-1$ that commutes with $H_0$, and $\mathcal{R}^{(n-1)}$ is of order $\geq n$. Writing the remainder $\mathcal{R}^{(n-1)}$ as a series of monomials of increasing order, $\mathcal{R}^{(n-1)} = \sum_{k=n}^\infty \epsilon^k R_k$, a Lie transform induced by a generating polynomial $\chi_n$ of order $n+2$ will result in the following form for the Hamiltonian:
\begin{equation}
(\mathcal{H} \circ \mathcal{T}_{n-1})\circ \phi_{\chi_n} = \epsilon^2 H_0 + Z^{(n-1)} + \epsilon^n (\{\chi_n, H_0\} + R_n) + \mathcal{O}(n+1).
\end{equation}
Then, $\chi_n$ is chosen such that $ \{\chi_n, H_0\} + R_n$ Poisson commutes with $H_0$; the remaining terms will be of higher orders from the construction. 

In particular, this implies that the Birkhoff normal form to 2nd order is obtained by choosing the 4th degree polynomial $\chi_2$ such that
$\{\chi_2, H_0\} + R_2$ commutes with $H_0$, where $R_2 \equiv P_2 + \{ {\chi_1}, P_1 \} + \{ {\chi_1}, \{ {\chi_1}, H_0 \} \}$.
For further details, including the method used to choose the functions $\chi_k$,  see \cite{Bambusi2014}, which includes a result about the time-scales at which the truncated system $H_0 + Z^{(n)}$ may be considered instead of the full system.

Following this recipe, we obtain the following truncated Birkhoff normal form of our system to order 4:
\begin{equation}
\begin{aligned}
\mathcal{H}^{\left(4\right)}&=\epsilon^{2}H_{0}+\epsilon^{4}Z_{2}+\epsilon^{6}Z_{4}
\\
H_{0}&=\sqrt{\frac{3}{2}}J_{2}+\sqrt{3}J_{3},
\\
Z_{2}&=-\frac{1}{64}\left(52J_{1}\left(J_{1}-J_{2}\right)\sin^{2}\psi_{1}+J_{2}\left(5J_{2}+6\sqrt{2}J_{3}\right)\right)
\\
Z_{4}&=a_{0}+J_{1}\left(a_{1}+b_{1}\cos2\psi_{1}\right)+J_{1}^{2}\left(a_{2}+b_{2}\cos2\psi_{1}\right) \\
&+J_{1}^{3}\left(a_{3}+b_{3}\cos2\psi_{1}\right),
\end{aligned}
\end{equation}
where 
$J_{1}=I_{1}$, $J_{2}=I_{1}+I_{2}$, $J_{3}=I_{3}$, $\psi_{1} = \phi_1 - \phi_2$, $\psi_2 = \phi_2$, $\psi_3 = \phi_3$, and 
 $\{ I_k, \phi_k \}$ are the action-angle variables associated with the linearized Hamiltonian $H_0$, $I_{k}=\frac{1}{\tau_s E_s}({\tilde w}_{k}^{2}+{\tilde p}_{k}^{2})$,
 and:
 \begin{equation}
 \begin{aligned}
a_0 &= \frac{4606 \sqrt{2} J_2^3-12401 J_2^2 J_3-38752 \sqrt{2} J_2 J_3^2+8736 J_3^3}{344064 \sqrt{3}},
\\
a_1 &= -\frac{J_2 \left(156017 \sqrt{2} J_2+1674 J_3\right)}{344064 \sqrt{3}},
\\
b_1 &= \frac{J_2 \left(837 J_3-43505 \sqrt{2} J_2\right)}{172032 \sqrt{3}},
\\
a_2 &= \frac{199522 \sqrt{6} J_2+837 \sqrt{3} J_3}{516096},
\\
b_2 &= \frac{124514 \sqrt{6} J_2-837 \sqrt{3} J_3}{516096},
\\
a_3 &= -\frac{45005}{28672 \sqrt{6}} \; , \; \; b_3 = -\frac{9001 \sqrt{\frac{3}{2}}}{28672}.
 \end{aligned}
 \end{equation}
\newpage

\bibliography{refregreg} 
\end{document}